\documentclass{elsart}
\usepackage{epsfig}
\usepackage{amsfonts}
\usepackage[verbose]{wrapfig}
\graphicspath{{/Users/alessandrocurioni/Private/Drafts/lxegrit/efficiency/oct06/NIMsub/}}
\newcommand{\g}{$\gamma$}

\newcommand{\na}{$^{22}$Na}

\newcommand{\yt}{$^{88}$Y}

\newcommand{\bc}{\begin{center}}
\newcommand{\ec}{\end{center}}
\newcommand{\be}{\begin{equation}}
\newcommand{\ee}{\end{equation}}
\newcommand{\bfg}{\begin{figure}}
\newcommand{\efg}{\end{figure}}
\newcommand{\bi}{\begin{itemize}}
\newcommand{\ei}{\end{itemize}}
\newcommand{\bt}{\begin{table}}
\newcommand{\enta}{\end{table}}
\newcommand{\keV}{\mbox{ke\hspace{-0.1em}V}}
\newcommand{\MeV}{\mbox{Me\hspace{-0.1em}V}}

\newcommand{\phibar}{\ensuremath{\bar{\varphi}}}

\renewcommand{\deg}{\ensuremath{^\circ}}

\begin{document}
\begin{frontmatter}

\title{A Study of the LXeGRIT Detection Efficiency for MeV Gamma-Rays
during the 2000 Balloon Flight Campaign}  
\author{A.~Curioni$^a$$^{,b}$, E.~Aprile$^a$, T. ~Doke $^c$, K.L.~Giboni$^a$,} 
\author{M.~Kobayashi$^a$$^{,c}$, U.G.~Oberlack$^d$}   
\address{$^a$Columbia Astrophysics Laboratory, Columbia University,
  New York, NY, USA}  
\address{$^b$Yale University, New Haven, CT, USA}
\address{$^c$Waseda University,Tokyo, Japan}
\address{$^d$Dept. of Physics \& Astronomy, Rice University, Houston,
  TX, USA} 

\begin{abstract}
   
LXeGRIT -- Liquid Xenon Gamma-Ray Imaging Telescope -- is the first
prototype of a Compton telescope for \MeV\ \g-ray astrophysics based
on a LXe time projection chamber.  
One of the most relevant figures of merit for a Compton telescope is
the detection efficiency for \g-rays, which depends on diverse
contributions such as detector geometry and passive materials,
trigger efficiency, dead time, etc. 
A detailed study of the efficiency of the LXeGRIT instrument, based
both on laboratory measurements and Monte Carlo simulations, is
presented in this paper.

\end{abstract}

\end{frontmatter}

\section{Introduction}

The concept of Compton telescope (CT) has proved the most successful
so far in imaging astrophysical \g-ray sources in the energy band
1-30~\MeV\ \cite{VSchon:book}. 
In a classical CT, without electron tracking, the initial direction of
the \g-ray is determined to within a circle of angular radius \phibar\
around the direction of the scattered \g-ray, given by the spatial
coordinates ($x_1, ~y_1,~z_1$) and ($x_2, ~y_2, ~z_2$), the subscripts
1 and 2 indicating the first and second interaction of the \g-ray.  
The scatter angle \phibar\ is given by an energy measurement, through
the Compton formula for photons on free electrons
\begin{equation}
E_{\gamma}' ~=~ \frac{E_{\gamma} \cdot m_e c^2}{E_{\gamma}\cdot (1 -
  cos\phibar) + m_e c^2} 
\end{equation}
where $E_{\gamma}'$ is the energy of the scattered \g-ray,
$E_{\gamma}$ the initial energy of the \g-ray, $m_e c^2$ the mass
of the electron (0.511~\MeV).
Any detector able to measure ($x_1 , ~y_1 , ~z_1, ~E_1$) and ($x_2,
~y_2, ~z_2, ~E_{\gamma} - E_1 $) for each \g-ray with sufficient
precision is a CT. The correct sequence of the two interactions must
also be known.     

\begin{figure}[htb]
\centering
\includegraphics[width=0.5\linewidth,clip]{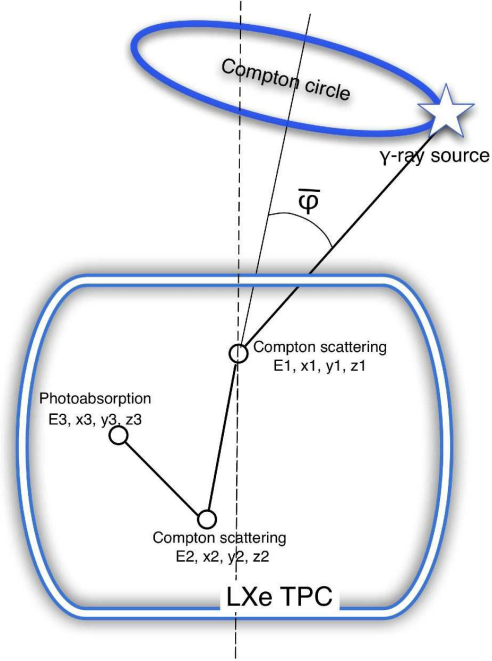}
\caption{Compton imaging in a single homogeneous detector, e.g. a
  LXeTPC. In this example the \g-ray Compton-scatters twice before
  being photo-absorbed; the separation between successive interactions
  in liquid Xe is typically few cm. Energy and position are
  measured for each interaction, therefore providing ($x_1 , ~y_1 ,
  ~z_1, ~E_1$) and ($x_2 , ~y_2, ~z_2, ~E_{\gamma} - E_1 $). As in a
  classical Compton telescope, the scatter angle \phibar\ is then
  determined from the Compton formula (eqn.~1) and the position of the
  \g-ray source is constrained within a Compton circle of angular
  radius \phibar\ around the direction of the scattered \g-ray, as
  determined by the two points ($x_1, ~y_1,~z_1$) and ($x_2, ~y_2,
  ~z_2$).}  
\label{f:compton}
\end{figure}

COMPTEL \cite{VSchon:93} on the Compton Gamma Ray Observatory (CGRO)
has been the most successful CT to date. COMPTEL was a double-scatter
CT with a liquid scintillator detector as \g-ray converter and a NaI
calorimeter. The two detectors were separated by a distance of 1.5~m
to allow a measurement of the scatter direction with an accuracy of
few degrees.
In laboratory calibration, COMPTEL obtained an effective area of
$\sim$20~cm$^2$ or less, out of a geometrical area of the upper
detector of 4188~cm$^2$, therefore with an efficiency well below
1$\%$. The sequencing of Compton interactions was based on the {\it
  time of flight} measurement between the upper and lower detector. In
flight the lower energy threshold for Compton imaging was 0.75~\MeV,
determined by rate considerations.   
An approach to CT largely different than COMPTEL was proposed some
time ago \cite{EAprile:93}, using a compact homogeneous detector with
greatly enhanced position resolution (the principle is schematically
shown in Fig.~\ref{f:compton}), specifically a liquid xenon time
projection chamber (LXeTPC), in order to increase the efficiency and
improve the background rejection capability well beyond the COMPTEL
achievement.    
The practical realization of this proposal has been the liquid xenon
\g-ray imaging telescope (LXeGRIT)
\cite{morelxegrit,EAprile:98.nim}. It has been tested at balloon
altitude four times: twice in 1997, mainly engineering balloon
flights, in 1999 with a heavy \g-ray shield and in 2000 with the
un-shielded LXeTPC. The year 2000-balloon flight, in particular,
lasted 27 hours and provided a data sample large enough to address all
the main technical issues, to give a thorough in-flight calibration
and to study the background in the near space environment 
\cite{ACurioni:2002,ACurioni:2004:PhDthesis}.   
%



Aim of this paper is to present a detailed study of the efficiency
of LXeGRIT to \MeV\ \g-rays, based both on experimental data and Monte
Carlo (MC) simulations. The focus will be on data taken just prior to
the 2000 flight.   
The overall detection efficiency results from the combination of
several contributions: detector geometry and passive materials,
trigger efficiency, dead  time and limited speed of the data
acquisition system, on-line and off-line selections. These different 
contributions will be analyzed separately.  
LXeGRIT will be dealt with as an imaging calorimeter but we will stop
short of discussing the key issue of imaging \MeV\ \g-ray sources, too
vast a subject to be included in the present work
(see, e.g., \cite{ACurioni:2004:PhDthesis}--Ch.~4).
The paper is organized as follows:
\begin{itemize}
\item [a. ] experimental measurements of the trigger efficiency,
  given in Sec.~\ref{sec:exp};
\item [b. ] MC simulation of LXeGRIT, in Sec.~\ref{sec:MC};   
\item [c. ] the full data/MC comparison, given in Sec.~\ref{sec:exa}.  
\end{itemize}
%


\section{\label{sec:exp} LXeGRIT Trigger Efficiency Measurement}

\begin{figure}[htb]
\centering
\includegraphics[width=0.9\linewidth,clip]{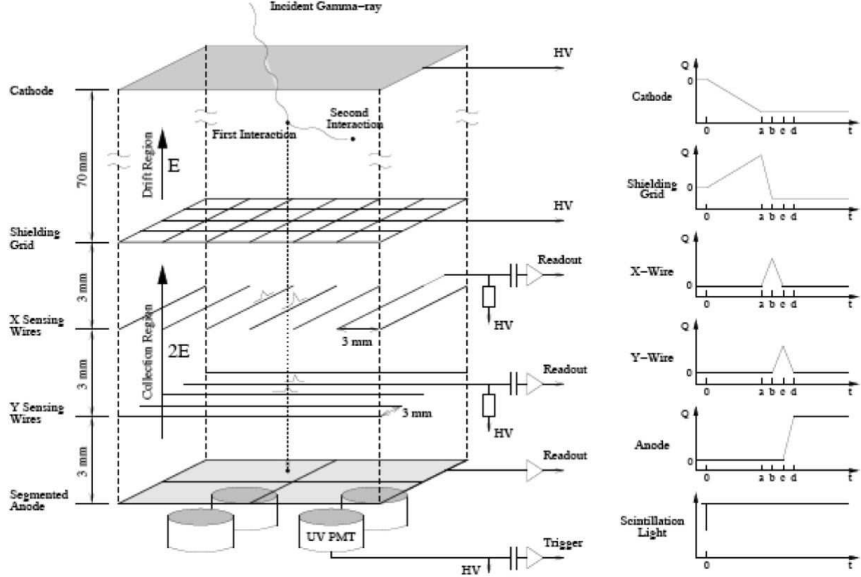}
\caption{Schematic of the LXeTPC readout structure (not to scale)
  with corresponding light trigger and charge pulse shapes. Adapted
  from \cite{EAprile:98.nim}.}  
\label{f:TPC_schematic}
\end{figure}

The LXeGRIT TPC is a position sensitive liquid ionization chamber
self-triggered by the fast xenon scintillation.  
When a \g-ray interacts in the fiducial volume, both scintillation
light and ionization charge are produced efficiently, with W-values of 
$W_\mathrm{ph}=$ 24~eV \cite{doke:W_ph} and $W_\mathrm{e}$ = 15.6~eV 
\cite{takahashi:W_e}. 
The VUV photons (178~nm) are detected by four photomultiplier tubes
(PMTs), which provide the event trigger and the initial time,
t$_0$.   
The ionization electrons drift under an external electric field,
inducing a signal on two parallel wire planes, screened by a Fritsch
grid. There are 62 wires in each plane and the pitch of the wires is
3~mm. The location of the hit wire(s) in the two wire planes provide
the $x$ and $y$ coordinates in the TPC reference frame, while the
time, measured starting from t$_0$, gives the interaction depth ($z$
coordinate). The wires do not collect the charge, which is finally
collected by one of four independent anodes, and the amplitude
measures the energy deposited in the interaction.    
Once the event has been built by the data acquisition system (DAQ),
on-line selections are applied to the signals from the wires and the
anodes. If not rejected, the \emph{raw event}, i.e. anode and wire
digitized waveforms, is written to disk.     
A detailed description of the LXeTPC and its working principle is
given in \cite{ACurioni:2004:PhDthesis,EAprile:98.nim}; a schematic
diagram of the LXeTPC showing the principle of operation, readout
structure and expected pulse shapes is given in
Fig.~\ref{f:TPC_schematic}.        




\begin{figure}[htb]
\centering
\includegraphics[bbllx=70,bblly=100,bburx=690,bbury=590,
  width=\linewidth,clip]{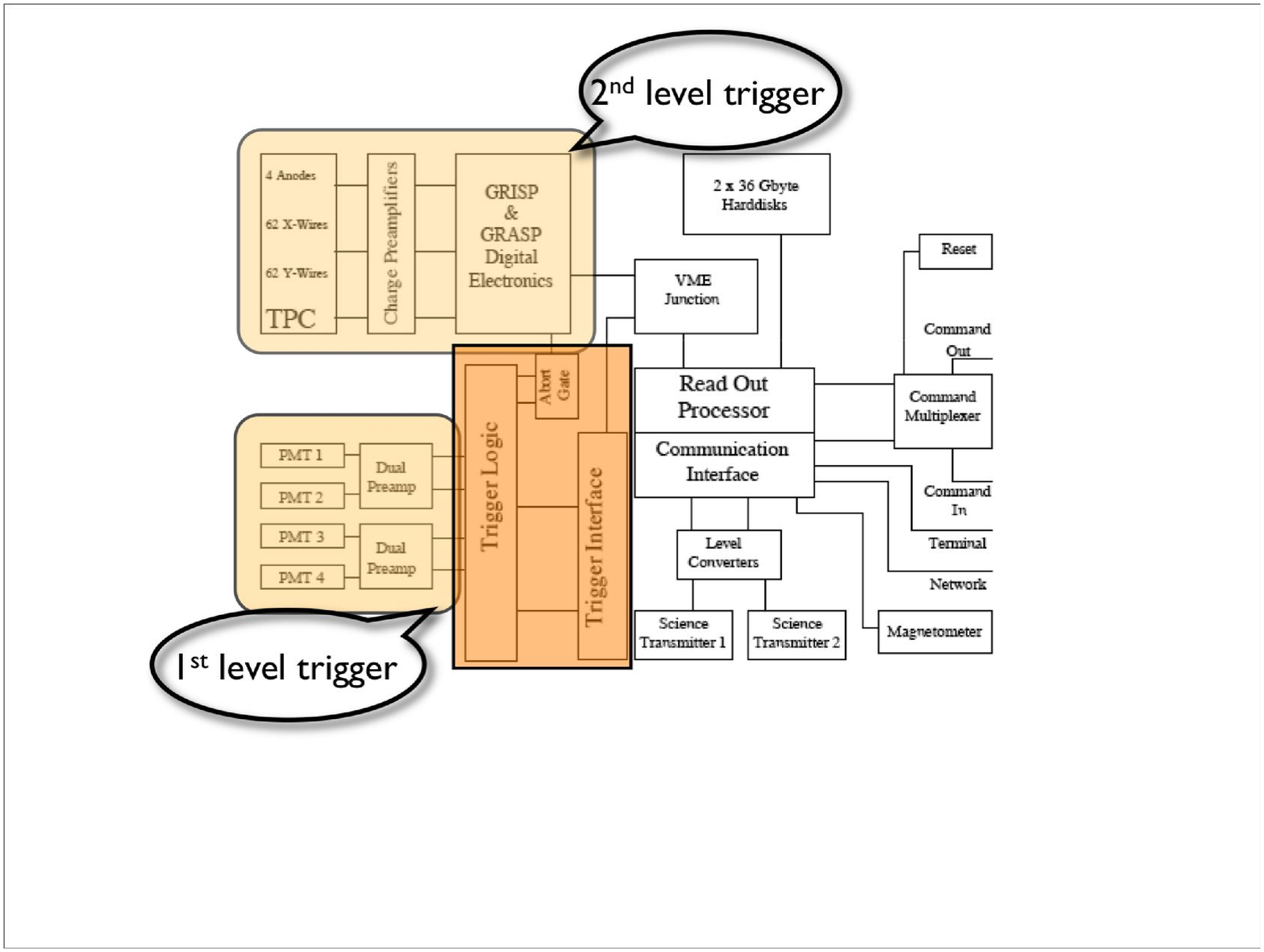}    
\caption{Diagram of the interconnections of the trigger and readout
  electronics of LXeGRIT. Adapted from \cite{EAprile:2000.ieee}.} 
\label{f:trigger_schematic}
\end{figure}

The event trigger works on two different levels: a {\it first level
  trigger} (FLT) which requires a signal from the PMTs and a {\it
  second level trigger} (SLT) which performs further selections based
on the anode/wire signature. The interconnections of the LXeGRIT
  trigger system and readout electronics are shown schematically in
  Fig.~\ref{f:trigger_schematic}.  


\subsection{\label{susec:LTE} First Level Trigger}

The FLT is provided by the logical OR of the four PMTs (Electron Tubes
9813QA). These 2-inch-diameter tubes, with a quantum efficiency of
about 15$\%$ at 178~nm, see the LXe volume through four 2.4 inch
diameter quartz windows with good transparency in the UV (88\% at
normal incidence). The bottom of the fiducial volume and the quartz
windows are separated by a 3~cm thick layer of LXe plus the collection
region, about 1~cm thick. This large separation reduces the solid
angle for events in the fiducial volume, and consequently the light
yield. Additional light losses are due to total reflection at the
window/PMT interface.  
Several photoelectrons are expected for every \MeV\ of energy
deposited in the fiducial volume \cite{EAprile:2002.ieee}. 
The FLT, while allowing a fast ($\sim$$\mu$s) decision, is unable to
select specific event categories of interest for Compton imaging.   
The 1999 balloon flight
\cite{EAprile:2000.spie,ACurioni:2004:PhDthesis} clearly demonstrated
that, when the energy spectrum of the incident \g-rays is soft, as for
the atmospheric \g-ray background, described by power law with
spectral index 2, a high FLT  efficiency is equivalent to accept a
dominant fraction of events at low energies which are of little if any
use for a CT, eventually \emph{reducing} the efficiency for Compton
imaging.       
For the 2000 flight, the light trigger efficiency was changed to
minimize the shortcomings detected during the previous flight. 

\begin{figure}[htb]
\centering
\includegraphics[bbllx=55,bblly=505,bburx=550,bbury=730,
	width=\linewidth,clip]{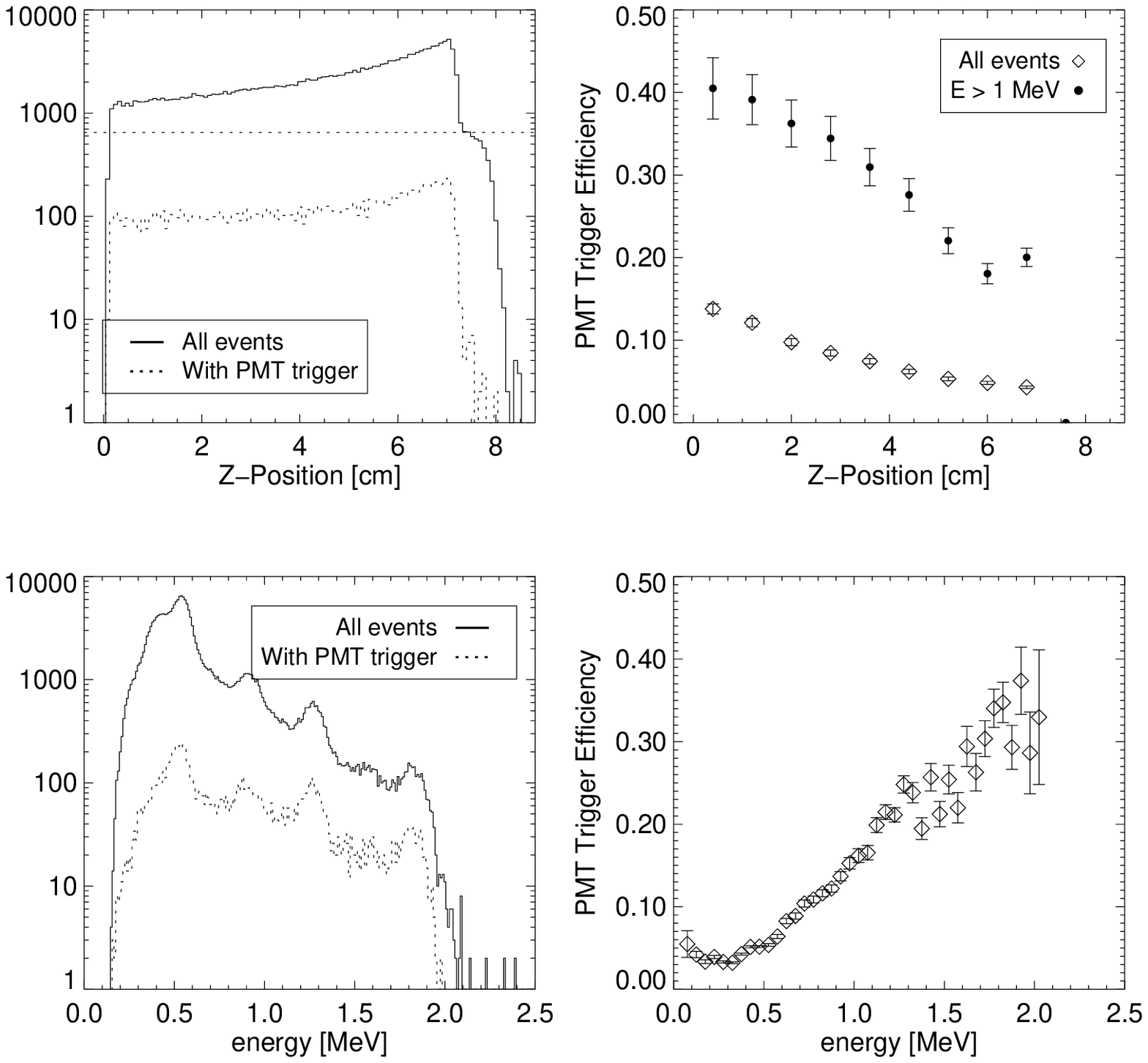}
\caption{$z$ dependence of the light trigger efficiency. {\it Left:}
  $z$ distribution for all events (continuous line) and events with a
  PMT trigger (dotted line). A sharp drop is visible at $z$~=~7~cm,
  corresponding to the maximum ``physical'' $z$. 
  Note the logarithmic scale. {\it Right:} Ratio of the two $z$
  distributions shown on the left, after correction for the fraction of
  random coincidence events -- {\it open diamonds}. The same ratio
  after selecting energy deposits larger than 1~\MeV -- {\it full
    circles}.}   
\label{f:LTE.zpos}
\end{figure}

\begin{figure}[htb]
\centering
\includegraphics[bbllx=55,bblly=275,bburx=550,bbury=500,
	width=\linewidth,clip]{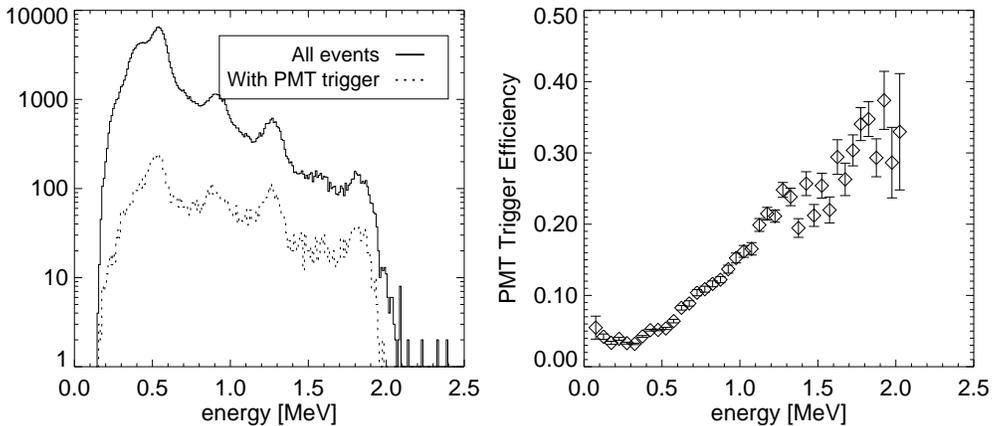}
\caption{Energy dependence of the light trigger efficiency. {\it
    Left:} Energy spectrum for \yt\ and \na\ source combined, for all
  events (continuous line) and events with a PMT trigger (dotted
  line). The 0.511, 0.898, 1.275 and 1.836 \MeV\ lines are clearly
  visible. {\it Right:} ratio of the two energy spectra, which gives
  the light trigger efficiency versus energy.}   
\label{f:LTE.energy}
\end{figure}

\begin{figure}[htb]
\centering
\includegraphics[bbllx=55,bblly=275,bburx=550,bbury=730,
	width=0.9\linewidth,clip]{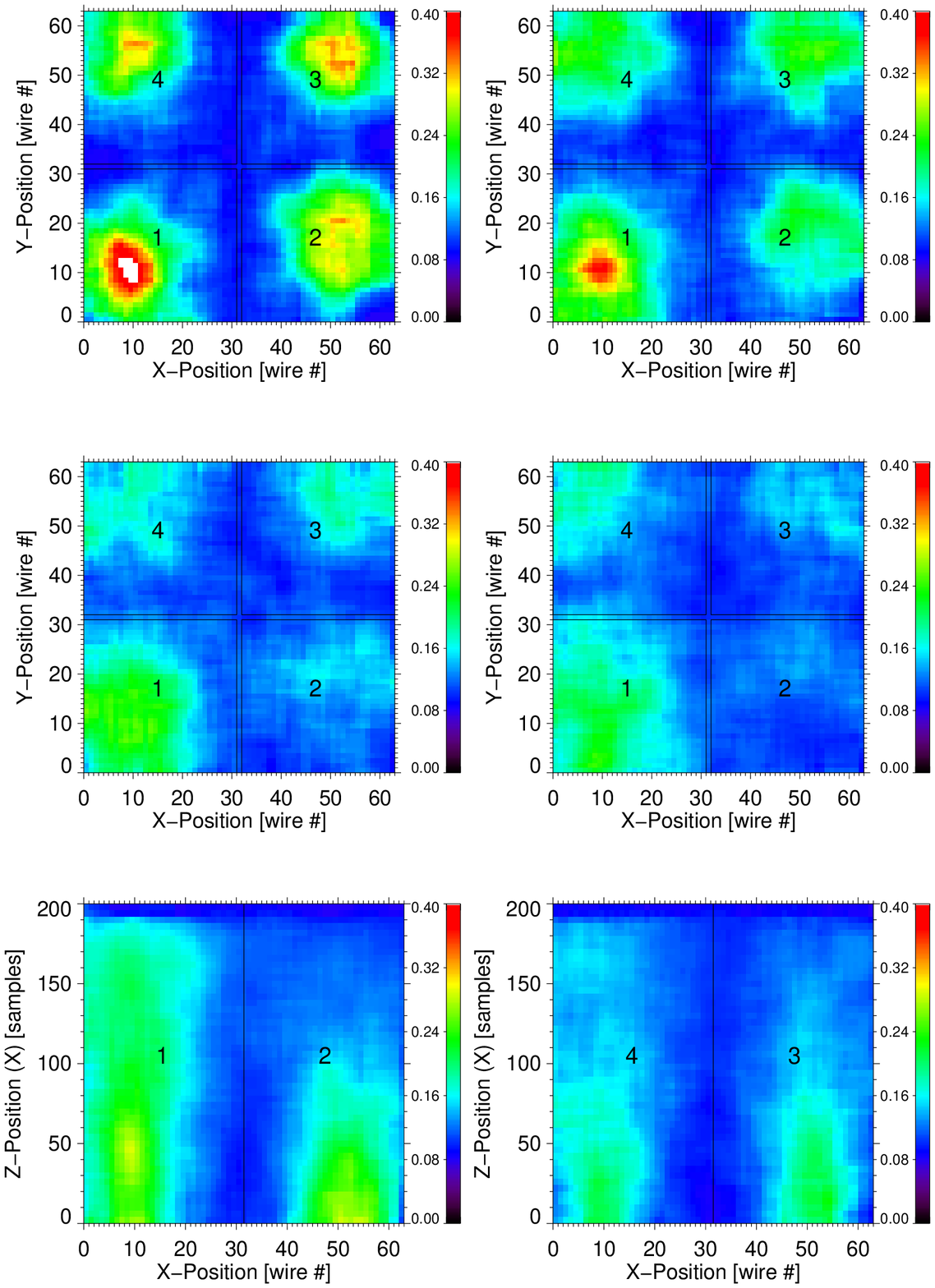}
\caption{Maps of light trigger efficiency in the $x-y$ plane, for four
  different $z$ slices, $\sim$1.7~cm thick. The four PMT locations
  are clearly visible as areas of enhanced efficiency. From the {\it 
    top-left} to the {\it bottom-right} panel we move from the $z$
  slice closest to the PMTs towards the cathode region.}
\label{f:LTE.xy}
\end{figure}

The efficiency was measured and resolved versus energy and position,
following the procedure described in \cite{UOberlack:2000.ieee}. 
The collected data fall into three classes:    
\begin{itemize}
\item[A. ] events with an interaction detected in the fiducial volume 
  (charge signal) and a light trigger which matches the external
  trigger -- this class is the one contributing to the {\it
    efficiency};    
\item[B. ] events with an interaction detected in the fiducial volume
  and a light trigger which does not match the external trigger (just
  a chance coincidence count) -- this class introduces a {\it
    background} in the measurement, that has to be accounted for;
\item[C. ] events with an interaction detected in the fiducial volume
  and without a light trigger -- this class gives the {\it
    inefficiency}. 
\end{itemize}
The efficiency is measured as the number of events in {\it class A}
divided by the total number of events subtracted {\it class B}.  
The measurement described in \cite{UOberlack:2000.ieee} used a
\na\ source, which emits three photons at the same time, one at
1.275~\MeV\ and two at 0.511~\MeV\ which are emitted back-to-back, so
that one of the two 0.511~\MeV\ photons could be used to tag the
source. The spatial correlation of the two photons allowed a low
contamination due to chance-coincidences. While this allowed a
detailed measurement of the spatial dependency of the trigger
efficiency for the 1999 balloon flight settings, the energy range of
that measurement was limited up to 0.511~\MeV, rather low for a
detector designed to work up to 10~\MeV.   
A similar measurement was repeated for the 2000 settings  using an
\yt\ source, which emits simultaneously two photons at 0.898 and
1.836~\MeV\ with little angular correlation, and, separately, a
\na\ source. The sources were loosely tagged detecting one of the
photons from the decay with a NaI(Tl) counter. Given a source rate of
several kHz and a low tagging efficiency, the level of contamination
due to chance coincidence counts was high, and, combined with the low
efficiency in the 2000 settings, more events ended up in {\it class B} 
rather than in {\it class A}. 
Nonetheless, the fraction of events in {\it class B} is independently
determined and the final result is then corrected for with good
accuracy. In fact, the $z$ (drift time) distribution for events
triggered by random coincidence has no physical meaning and extends
above the nominal cathode position (7~cm), which delimitates the TPC
active volume. This is visible in Fig.~\ref{f:LTE.zpos} where the $z$  
distribution drops sharply at $z$~=~7~cm. The shape of the $z$
distribution for events in {\it class B} was determined through data
where a random trigger was fed in as FLT. It did show no appreciable
dependence on $z$, so that the fraction of background events was read
out from the $z$ distribution itself.       
The efficiency of the FLT is in this way measured for energies up to
about 2~\MeV.  
The efficiency was resolved in a four dimensional data space --
energy, $x$, $y$, $z$ --  as made possible by the imaging capability
of the TPC. Since each event has to be spatially resolved and, at the
same time, the four PMTs see the total energy deposited in the
fiducial volume, the analysis was limited to 1-site events, for which
a univocal association between total energy loss and position is
given. The efficiency in the 4D space was described factoring out the
energy dependence, i.e.    
$$
\epsilon(E, x, y, z)~=~ \epsilon _1 (E) \times w (x, y, z) .
$$ 
where $w$ is a position dependent weighting factor.
The dependence of the efficiency on the deposited energy is shown in
Fig.~\ref{f:LTE.energy} for the four PMTs combined, using data from
both \na\ and \yt\ sources. A PMT mostly detects interactions
happening within its own quadrant and very few events are detected by
more than one PMT at the same time, because of the strong solid angle
effect on light collection. The response of each PMT is measured and
parameterized individually and the energy dependence is described
analytically.     
For use in MonteCarlo simulations, the spatial dependence was
parameterized in a lookup table with a granularity of 1.17~cm in $z$
and 1.8~cm in $x$ and $y$ (6 $z$ slices times 11$\times$11 bins in the
$x-y$ plane).    
The dependence on $z$ is shown in Fig.~\ref{f:LTE.zpos} for one PMT. As
expected, the efficiency decreases with increasing distance from the
PMT's location ($z$~=~-4~cm). The energy spectrum is the one in
Fig.~\ref{f:LTE.energy}; selecting energy deposits larger than
1~\MeV\, the overall efficiency is greatly increased.  
An example of $x-y$ efficiency map for the light trigger is shown
in Fig.~\ref{f:LTE.xy}, as measured using a tagged \yt\ source,
integrated over all energies. The four panels correspond to
four different $z$ slices, going from the bottom of the fiducial
volume ({\it upper-left}) to the top ({\it bottom-right}), i.e. moving
farther away from the PMTs. The impact of solid angle is apparent, and
closer to the PMTs the overall efficiency is higher and less
uniform. At locations corresponding to the four PMTs below the wire
structure an enhanced  efficiency is clearly visible. 


\subsection{\label{susec:SLT} Second Level Trigger}

\begin{figure}[htb]
\centering
\includegraphics[bbllx=55,bblly=505,bburx=550,bbury=730,
	width=\linewidth,clip]{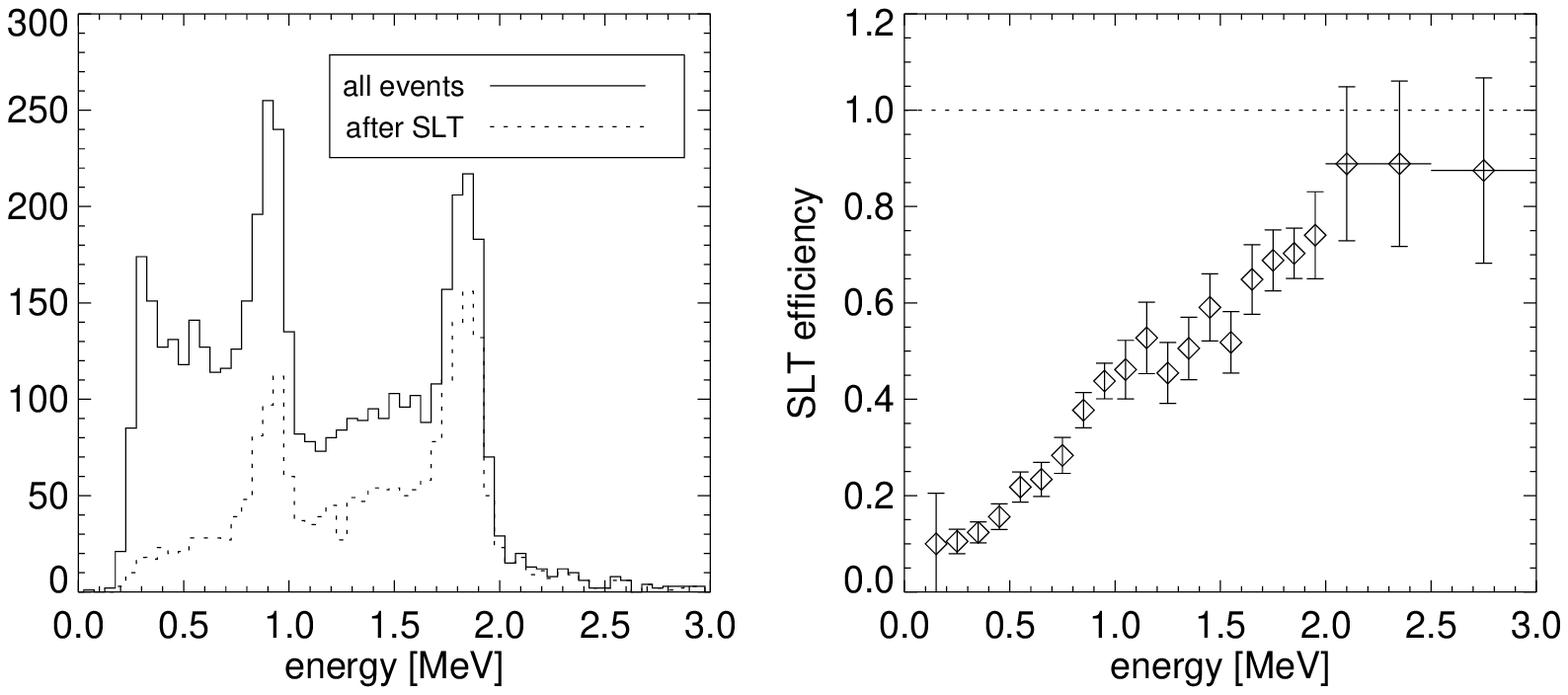}
\caption{Efficiency of the second level trigger (SLT). {\it Left:}
  energy spectra for an \yt\ source, combining all multiplicities,
  before (all events, {\it solid line}) and after SLT ({\it dashed
    line}). {\it Right:} ratio of the two energy spectra, which gives
  the SLT efficiency versuss. energy.} 
\label{f:minmax.1}
\end{figure}

\begin{figure}[htb]
\centering
\includegraphics[bbllx=30,bblly=555,bburx=550,bbury=750,
	width=\linewidth,clip]{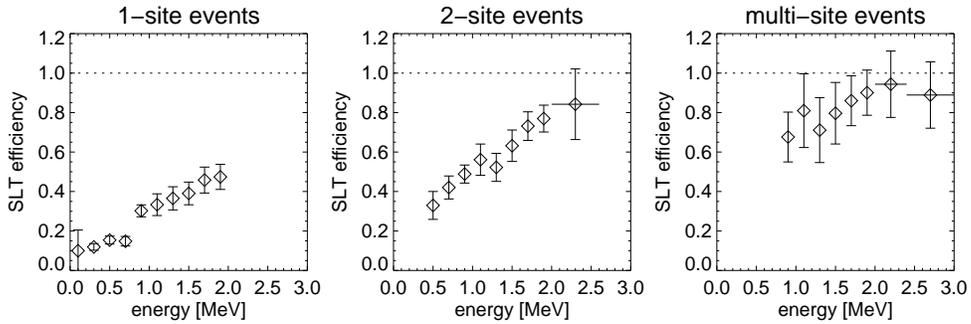}
\caption{Efficiency of the second level trigger for different
  multiplicities. {\it Left:} 1-site events; {\it middle:} 2-site
  events; {\it right:} multi-site events.} 
\label{f:minmax.2}
\end{figure}

The SLT is provided by on-line selections based on the wire and anode
signals: 
\begin{itemize}
\item[i.] the event is built (see Sec.~\ref{susec:DAQ}) and the number
  of threshold-crossings on the $x$ and $y$ wires ({\it wire hits}) is
  checked to be greater than a predefined minimum and less than a
  predefined maximum. This step requires little readout time, since
  each channel is 1~byte only. It is also sensitive to specific event
  multiplicities, relevant for a CT since Compton imaging requires at
  least two interactions; 
\item[ii.] the amplitude of the anode waveform is checked against a
  given threshold. Imposing a threshold on the anode waveform is very
  effective in rejecting low energy events which are useless for a CT,
  but, for LXeGRIT electronics, it requires to readout the digitized
  waveforms and the slowness of this process greatly reduces the
  usefulness of this on-line selection.   
\end{itemize}
The SLT has proven to be a powerful tool, although critically
dependent on the noise level of the wires, and requiring careful
monitoring.   
The impact of the SLT was verified for each experiment by applying
{\it a posteriori} the on-line selections to a sample of data acquired
with full digitization of all wires.
The outcome of one of these routine checks is shown in
Figs.~\ref{f:minmax.1}, \ref{f:minmax.2} for data taken with an
\yt\ source placed a few meters away from the TPC. The minimum number of
wire hits required both in the $x$ and in the $y$ wire plane was 6,
the maximum 16. Each interaction usually gives 1 or 2 hits, depending
on the energy deposit and the threshold of the specific wire. The
maximum has no impact on \g-events and mainly filters out relativistic 
charged particles crossing the chamber and giving many hits at the
same time. Ideally a minimum of 6 wire hits would select only events
with 3 or more interactions, but in a situation with non-negligible
electronic noise a certain fraction of 1-site events is accepted (see  
Fig.~\ref{f:minmax.2}). Moreover, the fraction of accepted events
depends on energy, since an interaction with large energy is more
likely to be detected over noise. For events with multiplicity 3 or
higher the SLT efficiency is $\sim 80\%$ above 1~\MeV, where most of
these events are.     
The main goal of this SLT filter is therefore to reject low energy (below
0.5~\MeV) 1-site events, especially important when the \g-ray flux has a
soft spectrum.
This simple procedure just counts the number of wire hits and does not
exploit the 3D imaging capability of the TPC. When data are analyzed
off-line, the $x$, $y$ and $z$ coordinates are matched and the
contamination due to noisy events is reduced to a truly negligible
level \cite{UOberlack:inprep}.    
%

\subsection{\label{susec:DAQ} Data acquisition }

The DAQ system developed for the 2000 flight of LXeGRIT is described
in \cite{EAprile:2000.ieee}. The total throughput of the system is
limited to about 1.6~MB/s, restricting the event building rate to
40-50 events/s when data are taken in a mode that reads all
information from anodes and wires, providing a full image of the
chamber (in this case the event size is about 30~kB).  
The more usual data-taking mode transfers only wire waveforms which
crossed preset thresholds together with the four anode waveforms and
the maximum rate of built events increases to 200-400~Hz, the actual
value heavily depending on the selection parameters and the specific
source. In flight conditions and for typical settings the average
event size is $\sim$5.5~kB.   
%
%

The deadtime behavior of the system is non-paralysable with respect to  
the DAQ, i.e. the rate at the trigger level $n$ and the recorded count
rate $m$ are related as \cite{r:knoll}
$$
 n = \frac{m}{1-m\tau} 
$$
where $\tau$ is the deadtime of the  system, $\sim$3~ms in 2000
flight; for $n \gg 1/\tau$, $ m \approx \frac{1}{\tau} \approx
0.3$~kHz.    
The performance of the DAQ is usually the main limiting factor to the
efficiency of LXeGRIT and imposes an upper limit to the maximum rate
of useful events, i.e. events written to disk and made available for
further analysis. Once this upper limit is saturated, the final
efficiency for Compton imaging can be improved only by being more
selective and enhancing the fraction of multi-site events in the
final data sample.
In comparison, the TPC itself, given a maximum charge collection time
of about 40~$\mu$s, is hardly dead-time limited in any realistic
situation.

\section{\label{sec:MC} Monte Carlo simulation of the LXeGRIT detector}

\subsection{\label{sec:descr} Mass model } 

The LXeGRIT TPC, with an active volume of a
18.6$\times$18.6$\times$7~cm$^3$ (2.4~l) is housed inside a stainless
steel cylindrical vessel, with a diameter of 35~cm and 11.5~cm high,
with an internal volume of about 10~l. The walls of the vessel are
3~mm thick, and the top flange is 10~mm thick, thinned to 5~mm in the
area covering the TPC.  
Between the bottom of the active volume and the bottom flange there is
a gap of 3.6~cm, filled with passive LXe.
To reduce passive LXe around the TPC electrodes structure,
stainless steel spacers are used on three sides (the fourth one 
housing the HV feedthrough).
Thermal insulation of the cold vessel is provided by a vacuum
cryostat. The lower section of the cryostat encloses the four PMTs
with their HV divider circuits and the electronics boards used to
decouple the HV bias to the wires from the signal coupling boards.
The cryostat, also made out of stainless steel, has a cylindrical
shape with 3~mm thick walls while and a 7~mm thick  top flange,
thinned to 5~mm above the sensitive area. The diameter of the vacuum
cryostat is 47.6~cm and the height is 36.1~cm. The total mass,
including LXe, is about 190~kg.   

\begin{figure}[htb]
\centering
\includegraphics[bbllx=220,bblly=210,bburx=440,bbury=500,
	width=0.42\linewidth,clip]{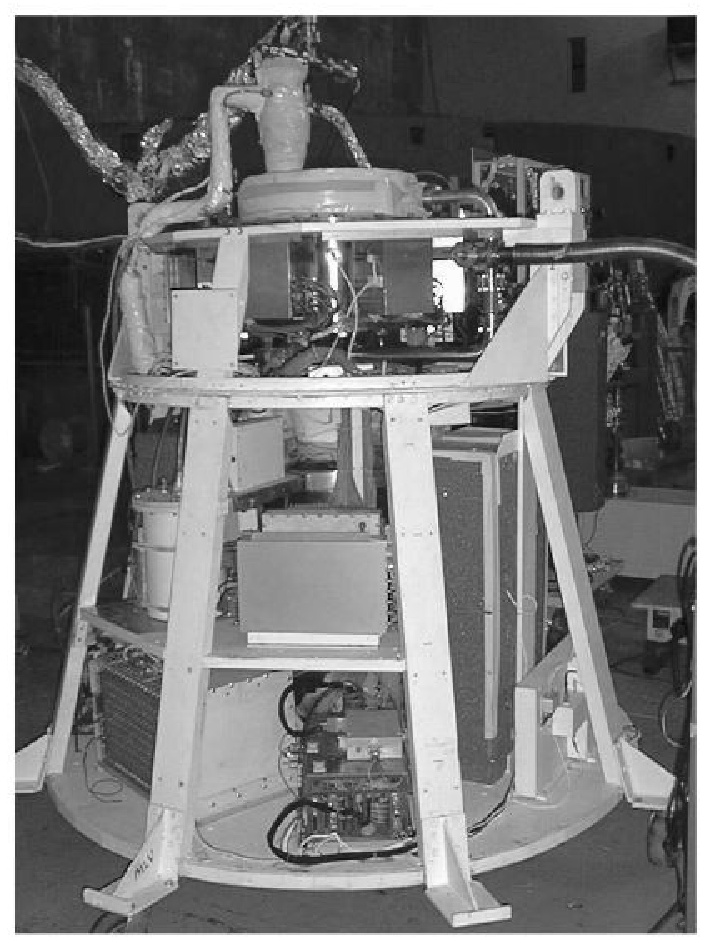}
\includegraphics[bbllx=20,bblly=210,bburx=600,bbury=730,width=0.57\linewidth,clip]{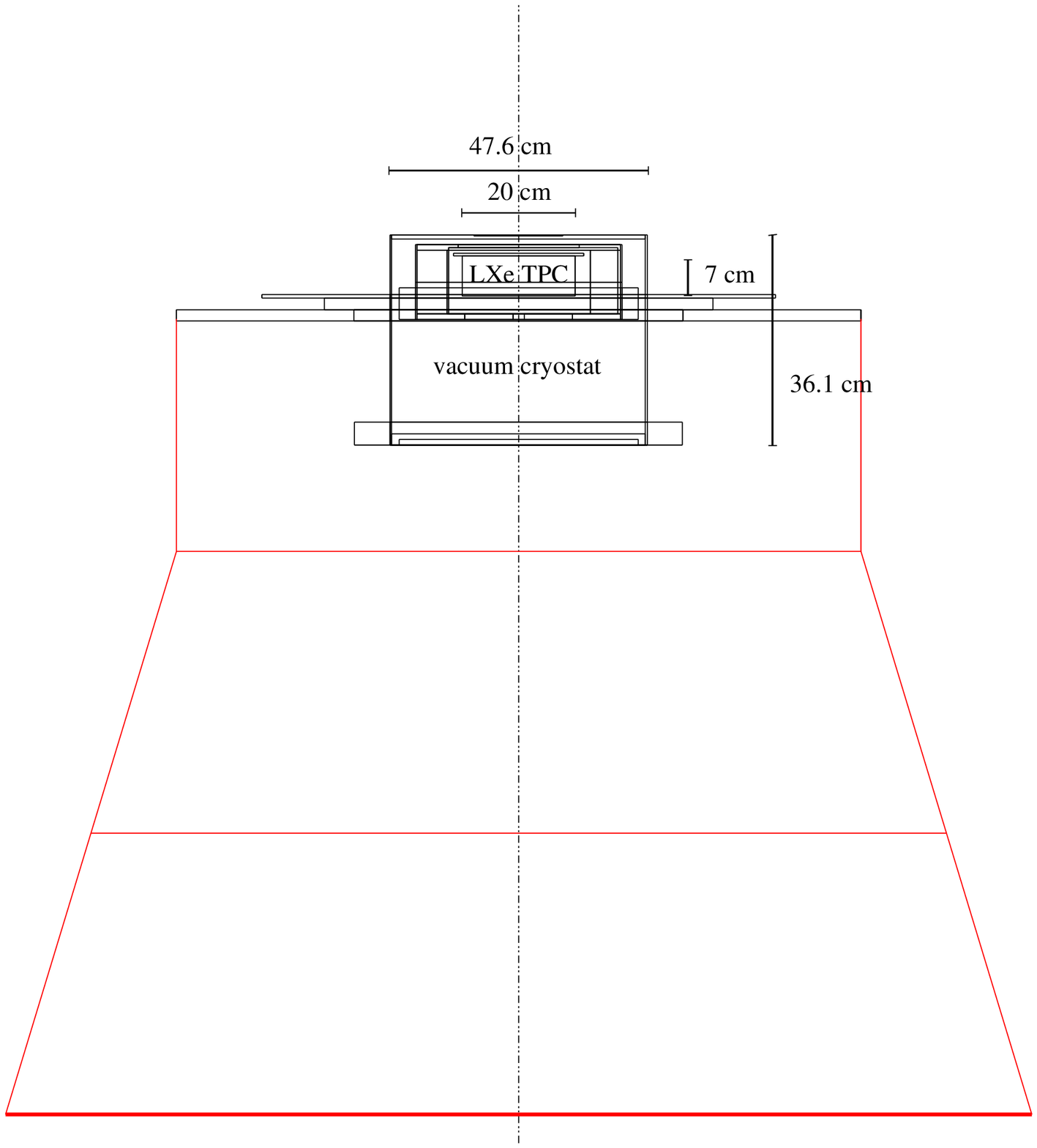}    
\caption{$Left:$ The LXeGRIT payload in the 2000 flight
 configuration. The upper section of the cryostat enclosing the LXeTPC
 is clearly visible; the lower section is hidden behind the boxes of
 the front-end electronics. On the lowest floor of the gondola the
 battery stack (right), the housekeeping computer (center), the VME
 crate of the digital electronics (left) are visible; on the
 intermediate shelfs, the trigger electronics box and the box with
 2$\times$36~GB data disks are visible. $Right:$ Schematic of the mass
 model of LXeGRIT instrument, in the $x-z$ plane. The LXeGRIT gondola
 has been modeled as if the mass was concentrated in three planes.}  
\label{f:gondola}
\end{figure}

A picture of the LXeGRIT payload in 2000 flight's gondola
configuration, is shown in Fig.~\ref{f:gondola}-$left$. The various
components of the equipment are rather evenly distributed and the most
abundant material is by far aluminum. 
Including the gondola in the MC mass model is necessary when a large
fraction of the \g-ray flux comes from below the LXeTPC, as it is the
case for the atmospheric \g-ray flux \cite{ACurioni:2002}.  
The gondola is modelled as a truncated cone, 102~cm high with a
diameter of 183~cm at the bottom and a diameter of 122~cm at the 
top(see Fig.~\ref{f:gondola}-$right$). 
A simplified mass model has been implemented, using parallel ``disks
and donuts'' at four different locations in $z$.
Its main features are shown in Fig.~\ref{f:gondola}-$right$. 
The accuracy of this model is expected to be adequate when the \g-ray
flux is azimuthally symmetric, or any azimuthal dependence is washed
out by the spinning of the instrument around the $z$ axis.

MC studies of the performance of the LXeGRIT instrument have been
based on the GEANT 3.21 detector simulation package \cite{r:geant}.  
%
%
Different \g-ray sources (from internal background to point-like
monochromatic sources to diffuse sources with power law spectra) have
been encoded in the simulation and are selectable through an input
card file. 
For each interaction of a \g-ray tracked through the mass model,
$x$,~$y$,~$z$, energy deposit and interaction type are recorded;
secondary Bremsstrahlung photons are tagged as such, regardless of
their specific interaction mechanism.  
For each photon which deposits at least 10~\keV\ in the fiducial
volume, an entry is created in the output file. 
Up to this point, only the mass model of the detector and the physics
of \g-ray interaction are accounted for. The other main effects are
then introduced in this order: first, treatment of the {\it charge
  signal}; second, treatment of the {\it light signal}. 

The response of the TPC to a {\it charge signal} is parameterized in
terms of minimum energy threshold and minimum spatial separation
required to resolve two close-by interactions.
Based on extensive comparisons of experimental and MC generated data
(\cite{ACurioni:2004:PhDthesis}--Ch.~2 and 3), the minimum separation in
the $x$ and $y$ coordinates is described by a normal distribution with
mean value 5~mm, 3~mm RMS and a sharp minimum of 4~mm; for the $z$
coordinate a normal distribution with mean value 4~mm, 1~mm RMS and a
sharp minimum of 3~mm is used. A pair of interactions must be resolved
{\it both in the} $x-z$ {\it and} $y-z$ planes, but can be confused in
$z$ {\it or} $x$ {\it or} $y$. If this condition is not fulfilled the
two interactions are clustered.  
The minimum energy threshold follows a normal distribution with
mean value 150~\keV, 40~\keV\ RMS and a sharp minimum of 40~\keV. This
energy threshold is applied to each single interaction (after
clustering) and not to the total energy loss in the fiducial
volume. The minimum energy threshold and minimum separation match the
performance of the TPC wire readout.
The deposited energy is then smeared according to the measured energy
resolution for the anode signals, $\Delta E / E = \sqrt{6.7\times
  10^{-3} \times E + 3.6\times 10^{-3}}$(FWHM), where $E$ is the
energy is MeV (\cite{ACurioni:2004:PhDthesis}--Ch.~3).    
This model mainly describes the {\it wire} response, which sets an
energy threshold higher than the anode threshold, together with
providing a finer spatial resolution. Only the energy resolution is
determined by the fitted {\it anode} signal.    

The efficiency of the light trigger is applied to  the MC data,
where each event is described as a sequence of localized energy
deposits ($E_i$, $x_i$, $y_i$, $z_i$), $i=1,...,n$, with $n$ the event 
multiplicity.  
The lookup tables described in Sec.~\ref{susec:LTE} give the 
probability to trigger the detector for each individual ($E_i$, $x_i$,
$y_i$, $z_i$). For 1-site events this immediately gives the trigger
probability for the entire event, while for multi-site events this is 
calculated combining the $n$ interactions. 
The corresponding statistical weight is then assigned to each \g-ray. 
 
The part of the detector response due to further on-line checks (SLT)
is introduced at this point of the data analysis, parameterized
versus energy for each multiplicity, as shown in Sec.~\ref{susec:SLT}.

\subsection{\label{susec:pa-ma-and-co} Event Multiplicity 
and Detection Efficiency}

\begin{figure}[htb]
\centering
\includegraphics[bbllx=55,bblly=405,bburx=550,bbury=730,
	width=0.8\linewidth,clip]{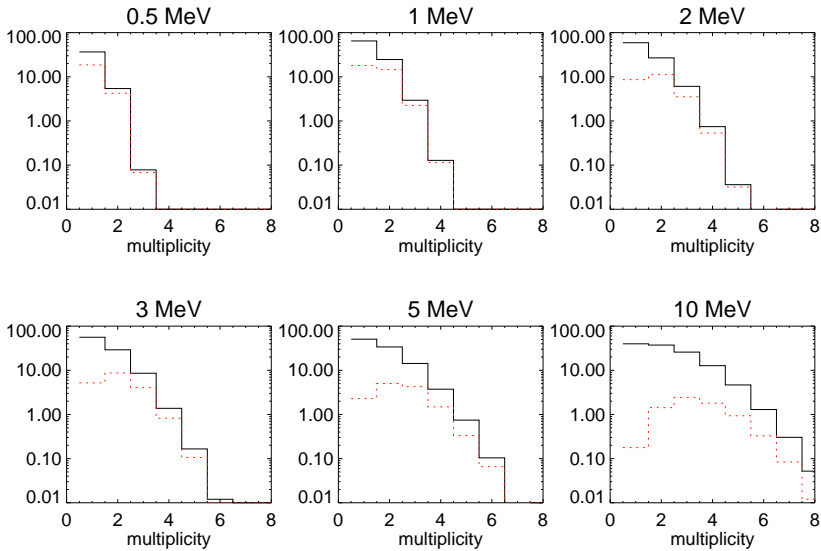}
\caption{Interaction multiplicity for different energies (0.5, 1, 2, 3, 5,
10~\MeV). The superimposed {\it dashed histogram} shows the interaction
multiplicity for fully contained events.}
\label{f:mult.1}
\end{figure}

The interaction multiplicity for 0.5, 1, 2, 3, 5 and 10~\MeV\ photons, 
for normal incidence and with the detector response as described
above, but without including the response of the FLT and SLT, DAQ
deadtime and off-line data analysis, is shown in Fig.~\ref{f:mult.1}.   
The same curves for fully contained events are superimposed. While the
single site events are the most numerous at all energies, the fully
contained events are more easily found in the multi site sample for
energies exceeding 1~\MeV. The interaction multiplicity is reduced
compared to the true number of interactions a \g-ray undergoes before
escaping the detector or being photoabsorbed. This reduction is due
both to the 150~\keV\ energy threshold and to the spatial confusion of
closeby interactions. While the energy threshold implies a neat loss
of detection efficiency, the effect of spatial confusion is to degrade 
the interaction multiplicity, even if the total energy is still
correctly measured. At energies above 5~\MeV\ the detected
multiplicity may substantially exceed the number of interactions of
the original photon, due to the presence of secondary Bremsstrahlung 
photons.   

\begin{figure}[htb]
\centering
\includegraphics[bbllx=70,bblly=510,bburx=330,bbury=730,
	width=0.5\linewidth,clip]{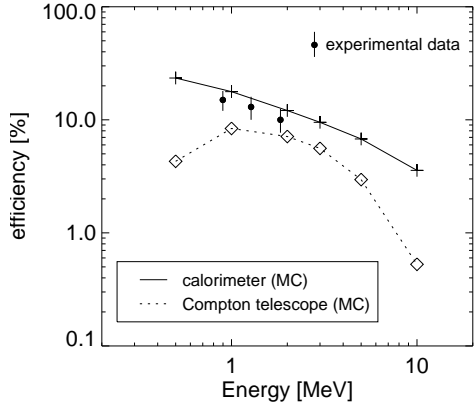}
\caption{Detection efficiency of the LXeTPC as a function of \g-ray
  energy. Two cases are considered, the {\it calorimeter mode} and the
  {\it Compton telescope mode}, which requires multiplicity 2 or
  larger and no pair-production or Bremsstrahlung interactions. The
  correct time sequence of interactions is assumed to be
  known. Experimental data are in {\it calorimeter mode} for 0.898 and
  1.836~\MeV\ (\yt) and 1.275 \MeV\ (\na), corrected for FLT, SLT and
  off-line efficiency, and DAQ deadtime; the details of the
  efficiency measurement are given in Sec.~\ref{sec:exa}. ($+$,
  calorimeter; $\lozenge$, Compton telescope. ) }   
\label{f:mult.4} 
\end{figure}

The detection efficiency calculated under the assumption of negligible
FLT, SLT and off-line reconstruction inefficiencies is shown in
Fig.~\ref{f:mult.4}, both for mere spectroscopy and Compton imaging.  
These results set an upper limit to the efficiency of the detector. As
a calorimeter, efficiencies for different multiplicities can be summed
up, while as a CT only efficiencies to double and multi site events
can be combined. As a CT, the efficiency is overestimated even for
negligible FLT and SLT inefficiencies, because multiple interactions
have to be time-sequenced before proceeding to Compton imaging.  

\begin{figure}[htb]
\centering
\includegraphics[bbllx=55,bblly=505,bburx=550,bbury=730,
	width=\linewidth,clip]{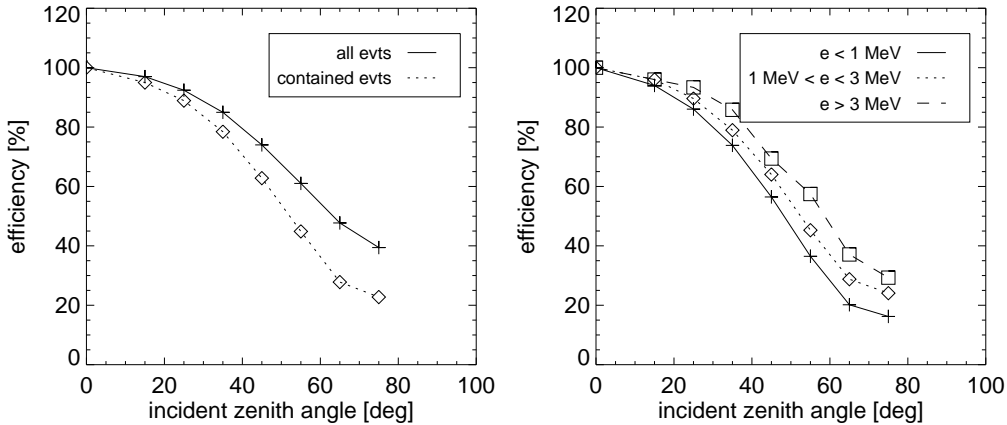}
\caption{Relative effective area as a function of the angular distance  
  from the zenith in detector coordinates, normalized for a source 
  on-axis (FoV). Photons have been generated following a power law
  spectrum with index 2. {\it Left:} Considering 
  all events (+ and continuous line) and fully contained events
  ($\lozenge$ and dotted line). {\it Right:} Dividing the fully
  contained events in energy bands: 0.85-1~\MeV\ (+ and continuous
  line), 1-3~\MeV\ ($\lozenge$ and dotted line) and 3-10~\MeV
  ($\square$ and dashed line).}    
\label{f:mult.5}
\end{figure}

In Fig.~\ref{f:mult.5} the relative effective area for sources at
different angles from the vertical ($z$ axis, in this case) is
shown, defining the field-of-view (FoV) of the instrument. A power
law spectrum with spectral index 2 was assumed, as is the case
for the Crab Nebula. The large FoV is about 100$^\circ$ opening
angle (FWHM) or 2.5~steradian. Two main parameters determine the FoV:
the  aspect ratio of detector diameter to depth and the amount of
passive materials  surrounding the detector. LXeGRIT has a large
aspect ratio of 18.6:7, which is the dominant contributor to the
decline in effective area with zenith angle. This geometrical effect
is energy-independent. Absorption in passive materials, especially
covering the side of the detector, yields an additional decline for
fully contained events (left panel) at large zenith angles. This
effect is energy-dependent, and most pronounced at low energies, as
shown in the right panel for contained events. The variation in
effective depth of the detector is a smaller energy-dependent effect.

\section{\label{sec:exa} Comparison of Experimental and Monte Carlo
  data} 

\begin{figure}[htb]
\centering
\includegraphics[bbllx=55,bblly=505,bburx=550,bbury=730,
	width=\linewidth,clip]{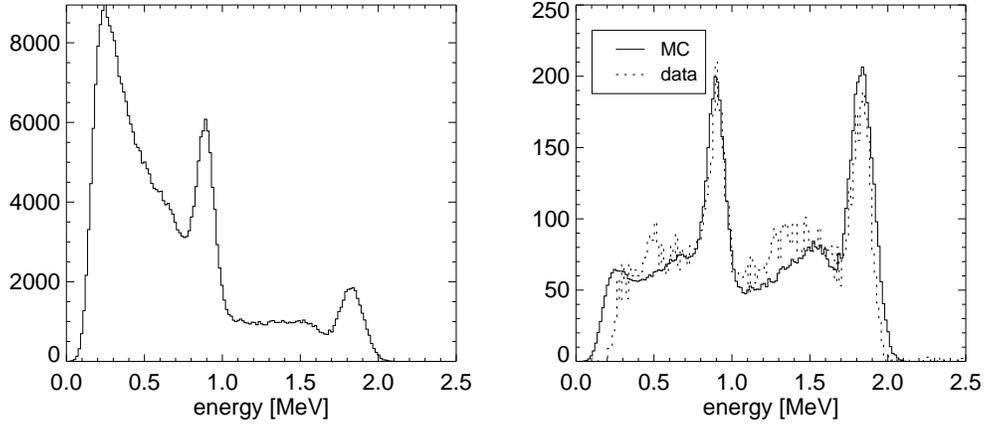}
\caption{ \emph{Left: } Energy spectrum for a MC simulated \yt\ source
  at distance, selecting 1-site events:; up to this point only the
  effects of a realistic detector geometry, passive materials, finite
  position resolution and an energy threshold of 150~\keV\ have been
  included. \emph{Right:} Energy spectrum for the same MC data after
  correcting for the efficiency at the first and second level
  trigger. The superimposed experimental data (\emph{dotted line})
  show a feature at 0.511~\MeV\ due to pair production events outside
  the fiducial volume. This feature is not reproduced in the MC energy
  spectrum, because this event class was excluded from the
  simulation. }    
\label{f:ytdi-MC.1}
\end{figure}

\begin{figure}[htb]
\centering
\includegraphics[bbllx=55,bblly=505,bburx=550,bbury=730,
	width=\linewidth,clip]{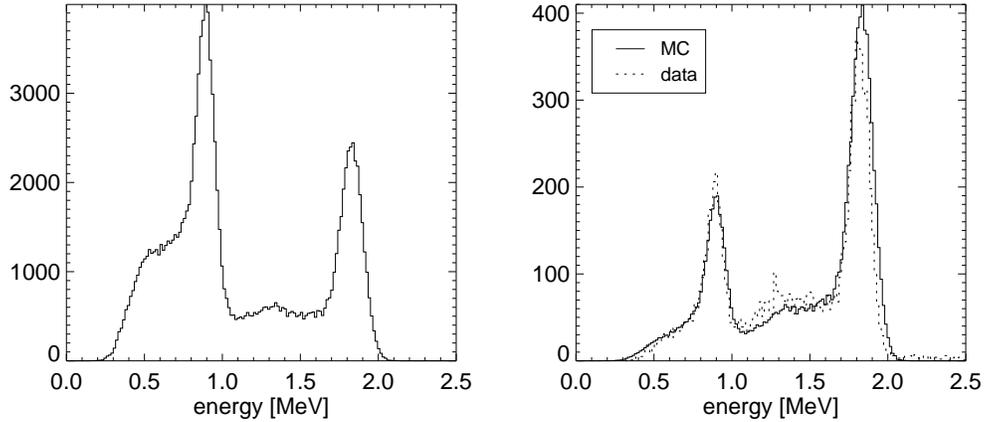}
\caption{ The same as Fig.~\ref{f:ytdi-MC.1}, but for 2-site
  events. The single escape peak is visible in the experimental
  data. } 
\label{f:ytdi-MC.2}
\end{figure}

\begin{figure}[htb]
\centering
\includegraphics[bbllx=70,bblly=505,bburx=330,bbury=730,
	width=0.5\linewidth,clip]{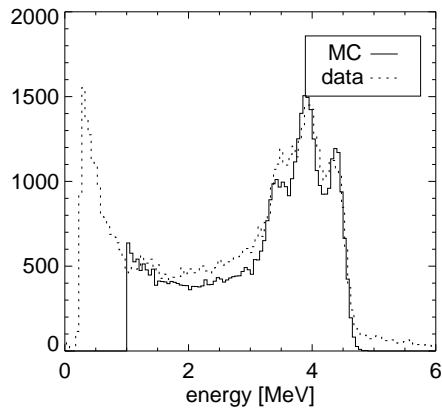}
\caption{Energy spectrum for a MC simulated Am-Be source (4.42~\MeV\
  \g-ray source) for 1-, 2-, 3-site events combined. The most
  prominent feature is the single-escape peak (3.92~\MeV). The full
  energy peak and the double escape peak are clearly detected,
  too. Superimposed ({\it dotted line}) are the experimental data for
  comparison.}  
\label{f:ambe}
\end{figure}

Figures~\ref{f:ytdi-MC.1}, \ref{f:ytdi-MC.2} and \ref{f:ambe} show
three examples of the accuracy of the Monte Carlo model in reproducing
experimental data.   
Fig.~\ref{f:ytdi-MC.1} shows the energy spectrum for 1-site events
from an \yt\ source placed at a 2~m distance from the TPC, 30\deg\
off-axis. On the left, the energy spectrum for MC data before
including FLT and SLT; on the right, the comparison of experimental
data and MC data after including FLT and SLT. A similar comparison for
2-site events from the same exposure is shown in
Fig.~\ref{f:ytdi-MC.2}. Notice that the we use an absolute
normalization, given the intensity of the source, which is the same
for the energy spectra in Fig.~\ref{f:ytdi-MC.1} and in
Fig.~\ref{f:ytdi-MC.2}. The overall suppression of 1-site events and 
the reduction of the large Compton continuum below 0.5~\MeV\ after FLT
and SLT is apparent in Fig.~\ref{f:ytdi-MC.1}. This is crucial for
LXeGRIT, since its efficiency is limited by the DAQ livetime fraction.    
Fig.~\ref{f:ambe} shows an Am-Be energy spectrum, combining 
multiplicities up to 3: since the cross-section  for pair-production 
is large at 4.4~\MeV, the single escape peak (3.92~\MeV) is the
dominant feature. The full energy peak (FEP) and the double escape
peak are also well identified.  
At energies above 4~\MeV\ Bremsstrahlung plays a non negligible role 
in LXe \footnote{The {\it critical energy} E$_c$, above which
Bremsstrahlung dominates ionization as an energy loss mechanism for
electrons, is 11~\MeV\ in LXe, but the energy spectrum is
significantly modified by Bremsstrahlung for energies as  
low as 4~\MeV.} and it contributes to the lack of sharpness in the 
energy peaks.   
MC simulated data have been superimposed and the comparison shows that
the main features are well reproduced. Neutrons from the source were
not included in this simulation \footnote{The intensity of the Am-Be
  source was unknown and the MC has been normalized to match the
  single escape peak.}.  
%

\subsection{\label{susec:exa} A detailed example of efficiency calculation}

A specific example of efficiency calculation for an \yt\ source at a
distance of 2~m, on-axis, is given in this section. \yt\ emits two
photons at 0.898 (branching ratio 94.4$\%$) and 1.836~\MeV\ (100$\%$).   
The total source rate was 2738~kBq, therefore the source was emitting
1.836~\MeV\ photons with a rate of 2738~kHz and 0.898~\MeV\ photons
with a rate of 2585~kHz. Taking into account a 5700~s exposure and
solid angle, the 18.6$\times$18.6~cm$^2$ geometrical area of the TPC
was hit by 1.08$\times$10$^7$ 1.836~\MeV\ photons and
1.01$\times$10$^7$ 0.898~\MeV\ photons.   
The DAQ livetime fraction was 50$\%$ livetime and the efficiency in
writing to disk was only 18$\%$, for only that fraction of the
bandwidth of disk writing was used. This last efficiency is usually
$\sim$100$\%$.  
Correcting for these two inefficiencies, one is left with
9.7$\times$10$^5$ 1.836~\MeV\ photons and  9.1$\times$10$^5$
0.898~\MeV\ photons.  
Counting the events in the two FEP gives the overall detection
efficiency. This procedure introduces an uncertainty of about 5$\%$
because of the necessary background subtraction. The results are: 
\begin{itemize}
\item 0.898~\MeV: 0.39$\times$10$^4$ events, i.e. 0.43$\%$ efficiency
\item 1.836~\MeV: 1.30$\times$10$^4$ events, i.e. 1.34$\%$ efficiency
\end{itemize}
These small figures require some explanations. As discussed in
Sec.~\ref{susec:pa-ma-and-co}, Fig.~\ref{f:mult.4} gives an upper
limit to the detection efficiency, including only the efficiency
for containment and the inefficiency due to passive materials. It is
about 18$\%$ at 0.898~\MeV\ and 11$\%$ at 1.836~\MeV. The light
trigger efficiency alone reduces this detection efficiency by a
factor of almost 9 at 0.898~\MeV\ and a factor of 3 at 1.836~\MeV. The
SLT further reduces the efficiency by a factor of 3 at 0.898~\MeV\ and
a factor 1.7 at 1.836~\MeV\ (Fig.~\ref{f:minmax.1}, from the same 
experiment). This decreases the efficiency to 0.6$\%$ at 0.898~\MeV\ and
1.9$\%$ at 1.836~\MeV. Losses due to noisy wires, events with
excessive baseline noise on the anodes, and other anomalous events
identified and rejected in the off-line analysis, account for the
missing part (\cite{ACurioni:2004:PhDthesis}--Ch.~3, and
\cite{UOberlack:inprep}).         
The efficiency differs for different multiplicities, mainly as a
result of the SLT, which is designed to enhance the fraction of
multi-site events. 
Inefficiencies due to DAQ livetime and disk writing are independent of 
the interaction multiplicity, and the FLT efficiency has on average
little dependence on the interaction multiplicity. 

There are various sources of uncertainty on the specific efficiencies,
some of them systematically lowering the final figure:
\begin{itemize}
\item the overall {\it measured} efficiency is known quite precisely,
  the main uncertainty coming from the subtraction of the background
  beneath the line ($\sim5$\%);   
\item the FLT efficiency is known with 5$\%$ precision for 1-site
  events, but assuming the same efficiency for higher multiplicities
  introduces a larger error. The  5$\%$ precision is restored for
  multi-site events through a complete MC simulation, as in
  Figs.~\ref{f:ytdi-MC.1}, \ref{f:ytdi-MC.2}, \ref{f:ambe};    
\item for a typical data set, the SLT efficiency is known within 5$\%$
  for all the multiplicities, statistics usually being the limiting factor;
\item the efficiency of the off-line procedure to extract the signal
  is more easily evaluated for 1-site events. In this case the
  efficiency in finding a good signal is very close to 100$\%$ for
  energies larger than few hundreds \keV, and the 30$\%$ inefficiency
  comes from the fraction of ``noisy'' events rejected off-line. For
  multi-site events the efficiency can be lower since, for
  instance, the chance of \emph{not} hitting a noisy wire decreases
  with interaction multiplicity.  
\item the precision of the MC expectation before FLT and SLT should be
  within 5$\%$ once the assumed conditions are matched. Many factors
  can systematically reduce the detection efficiency: noisy or dead
  wires, higher energy thresholds etc. Again, these factors are more
  relevant for higher multiplicities.   
\end{itemize}
It is clear from comparing the overall {\it expected} efficiencies to
the {\it measured} ones, as in Fig.~\ref{f:mult.4} (where the measured
efficiencies have been corrected for the estimated FLT, SLT and off-line
efficiencies), that the expectation is optimistic but within reason,
the discrepancy ranging between 15$\%$ and 30$\%$ (roughly equivalent,
for example, to the effect of 12 noisy wires out of 124), for
different energies and multiplicities.  
As a conclusion, it is easy at this point to extrapolate what the
efficiency would be if the DAQ could handle a 10 times larger
rate. With a light trigger efficiency as high as the one obtained in
1999 for the same detector (50\% or higher at 0.511~\MeV,
\cite{UOberlack:2000.ieee}) the final efficiency could be close to the
one depicted in Sec.~\ref{susec:pa-ma-and-co}.   

\section{Conclusions}

In this paper we have presented a study of the efficiency of the
LXeGRIT instrument for detecting  \MeV\ \g-rays. We have shown that
the detector response is well understood in its various components and
well reproduced through MC methods. 
The LXeGRIT DAQ clearly constitutes a severe bottleneck when dealing
with sources which generate a relatively high trigger rate (few kHz),
not an uncommon situation, given the size of the LXeTPC.  
For the 2000 balloon flight in year 2000, we chose to {\it reduce} the  
efficiency at the trigger level, to specifically select multiple
Compton events in the few-\MeV\ region. The rationale of this choice
was to to maximize the fraction of multiple Compton events given that
the maximum event rate was fixed by saturating the DAQ bandwidth. 
If not limited by the current DAQ, this same prototype could have
achieved a detection efficiency as a calorimeter close to 20$\%$ at
1~\MeV, which can be obtained in the current configuration only for
sources generating a trigger rate lower than 100~Hz.   

\section*{Acknowledgments}

This work was supported by NASA grant NAG5-5108 to the Columbia
Astrophysics Laboratory.


\end{document}